\documentclass[aps,prb,reprint,superscriptaddress]{revtex4-1}
\usepackage{amsmath}
\usepackage{multirow}
\usepackage{graphicx}
\usepackage{epstopdf}
\usepackage{tabularx, booktabs}
\usepackage{pbox}
\newcolumntype{Y}{>{\centering\arraybackslash}X}
\usepackage[usenames,dvipsnames]{color}
\usepackage[caption=false,position=b,singlelinecheck=off,font=normalsize,labelfont=bf,justification=justified]{subfig}
\usepackage{hyperref}

\hypersetup{
	colorlinks=true,
	linkcolor=blue,
	filecolor=black,      
	urlcolor=blue,
	citecolor=blue,
}
\graphicspath{{Figures_PDF/}}

\newcommand{\Ca}{Ca$_2$RuO$_4$}
\newcommand{\Cadoped}{Ca$_{2-x}$La$_x$RuO$_4$}

\begin{document}

\title{Tuning of the Ru$^{\mathbf{4+}}$ ground-state orbital population in the $\mathbf{4d^4}$ Mott insulator \Ca{} achieved by La doping} 

\author{D. Pincini}
\email{davide.pincini.14@ucl.ac.uk}
\affiliation{London Centre for Nanotechnology and Department of Physics and Astronomy, University College London, Gower Street, London WC1E 6BT, UK}
\affiliation{Diamond Light Source Ltd., Harwell Science \& Innovation Campus, Didcot, Oxfordshire OX11 0DE, UK}

\author{L.S.I. Veiga}
\affiliation{London Centre for Nanotechnology and Department of Physics and Astronomy, University College London, Gower Street, London WC1E 6BT, UK}

\author{C.D. Dashwood}
\affiliation{London Centre for Nanotechnology and Department of Physics and Astronomy, University College London, Gower Street, London WC1E 6BT, UK}

\author{F. Forte}
\affiliation{CNR-SPIN, IT-84084 Fisciano (SA), Italy}
\affiliation{Dipartimento di Fisica ``E. R. Caianiello", Universit\`a di Salerno, IT-84084 Fisciano (SA), Italy}

\author{M. Cuoco}
\affiliation{CNR-SPIN, IT-84084 Fisciano (SA), Italy}
\affiliation{Dipartimento di Fisica ``E. R. Caianiello", Universit\`a di Salerno, IT-84084 Fisciano (SA), Italy}

\author{R.S. Perry}
\affiliation{London Centre for Nanotechnology and Department of Physics and Astronomy, University College London, Gower Street, London WC1E 6BT, UK}

\author{P. Bencok}
\affiliation{Diamond Light Source Ltd., Harwell Science \& Innovation Campus, Didcot, Oxfordshire OX11 0DE, UK}

\author{A.T. Boothroyd}
\address{Clarendon Laboratory, Department of Physics, University of Oxford, Oxford OX1 3PU, UK}

\author{D.F. McMorrow}
\address{London Centre for Nanotechnology and Department of Physics and Astronomy, University College London, Gower Street, London WC1E 6BT, UK}

\date{\today}

\begin{abstract}
The ground-state orbital occupancy of the Ru$^{4+}$ ion in \Cadoped{} [$x = 0$, $0.05(1)$, $0.07(1)$ and $0.12(1)$] was investigated by performing X-ray absorption spectroscopy (XAS) in the vicinity of the O K edge as a function of angle between the incident beam and the surface of the single-crystal samples. A minimal model of the hybridization between the O $2p$ states probed at the K edge and the Ru $4d$ orbitals was used to analyze the XAS data, allowing the ratio of hole occupancies $n_{xy}/n_{yz,zx}$ to be determined as a function of doping and temperature. For the samples displaying a low-temperature insulating ground-state ($x\leq0.07$), $n_{xy}/n_{yz,zx}$ is found to increase significantly with increasing doping, with increasing temperature acting to further enhance $n_{xy}/n_{yz,zx}$. For the $x=0.12$ sample, which has a metallic ground-state, the XAS spectra are found to be independent of temperature, and not to be describable by the minimal hybridization model, while being qualitatively similar to the spectra displayed by the $x\leq0.07$ samples above their insulating to metallic transitions. To understand the origin of the evolution of the electronic structure of Ca$_{2-x}$La$_x$RuO$4$ across its phase diagram, we have performed theoretical calculations based on a model Hamiltonian, comprising electron-electron correlations, crystal field ($\Delta$) and spin-orbit coupling ($\lambda$), of a Ru-O-Ru cluster, with realistic values used to parameterize the various interactions taken from the literature. Our calculations of the Ru hole occupancy as a function of $\Delta/\lambda$ provide an excellent description of the general trends displayed by the data. In particular they establish that the enhancement of $n_{xy}/n_{yz,zx}$ is driven by significant modifications to the crystal field as the tetragonal distortion of the RuO$_6$ octahedral changes from compressive to tensile with La doping. We have also used our model to show that the hole occupancy of the O $2p$ and Ru $4d$ orbitals display the same general trend as a function of $\Delta/\lambda$, thus validating the minimal hybridization model used to analyze the data. In essence, our results suggest that the predominant mechanism driving the emergence of the low-temperature metallic phase in La doped \Ca\ is the  structurally induced redistribution of holes within the $t_{2g}$ orbitals, rather that the injection of free carriers.

\end{abstract}

\maketitle

\section{Introduction}

\Ca{} has attracted considerable attention in recent years as the Mott-insulating analogue of the unconventional superconductor Sr$_2$RuO$_4$ \cite{Cao1997,nakatsuji_ca_1997,Braden1998a,Alexander1999,mizokawa_spin-orbit_2001,Liebsch2007,Gorelov2010b,zegkinoglou_orbital_2005,khaliullin_excitonic_2013,fatuzzo_spin-orbit-induced_2015,jain_higgs_2017,Kunkemoller2015a,
Kunkemoller2017,Sutter2017a,Das2018,Pincini2018a}. A well-documented metal to insulator transition (MIT) occurs at $T_\textrm{\tiny MIT}=357$~K, concomitant with a first-order structural transition from a high-temperature quasi-tetragonal phase to a low-temperature orthorhombic one ($Pbca$ space group)\cite{Alexander1999,fukazawa_filling_2001,Cao1997,nakatsuji_ca_1997}. Contrary to many Mott insulators, such as La$_2$CuO$_4$, the nature of the insulating ground state cannot be accounted for by a simple half-filled single-band scenario where the Mott gap emerges due to a high Coulomb interaction ($U$) to bandwidth ($W$) ratio. In the case of the $2/3$-filled $t_{2g}$ manifold of the Ru$^{4+}$ ion ($4d^4$) in \Ca{}, the electronic band structure is strongly affected by Hund's and spin-orbit couplings, as well as by the crystal field\cite{Liebsch2007,Gorelov2010b,Georges2012a,Sutter2017a,Das2018}.

Angle-resolved photoemission measurements \cite{Sutter2017a} and \textit{ab-initio} calculations \cite{Liebsch2007,Gorelov2010b} indicated that the Mott insulating state of \Ca{} is triggered by the stabilization of the $xy$ orbitals induced by tetragonal compression of the RuO$_6$ octahedra in the low temperature $Pbca$ structure \cite{Braden1998a,Pincini2018a}: in this picture, a Mott gap is opened by Coulomb interactions in the narrower half-filled bands spanned by the $yz,zx$ orbitals (for which $W<U$), with a lower band of $xy$ character. In general, the low-energy multiplet structure of the Ru$^{4+}$ ion is expected to arise from the competition between the tetragonal crystal field $\Delta$ and the sizeable spin-orbit coupling (SOC) $\lambda$ of $4d$ electrons\cite{Nakatsuji2000d,Liu2011a,Liu2013,Kurokawa2002,mizokawa_spin-orbit_2001,fatuzzo_spin-orbit-induced_2015,Das2018}. The latter acts to weakly reconstruct the $xy$, $yz$ and $zx$ orbital configuration yielding a ground-state wave function of mixed orbital character. Although SOC plays an important role in determining the nature of the low-energy spin and orbital excitations in \Ca{}, previous studies\cite{Sutter2017a} have suggested that it has only a marginal role in driving the insulating phase of this system. This is because the SOC 
is not large enough to render a strongly coupled spin-orbit  $J_\textrm{eff}$ state, as is the case for certain iridates\cite{kim_novel_2008}. Moreover, even if it were, due to the $d^4$ filling of the Ru orbitals, the strong SOC regime would result in a singlet $J_\textrm{eff}=0$ local ground state, which would not facilitate the opening of a Mott gap.

The coupling between lattice and orbital degrees of freedom was highlighted by recent X-ray near-edge absorption spectroscopy (XANES) studies\cite{mizokawa_spin-orbit_2001,fatuzzo_spin-orbit-induced_2015,Das2018}, which revealed that the ground-state $t_{2g}$ orbital population can be effectively tuned by changing the ratio $\Delta/\lambda$, analogous to the case of the Ir$^{4+}$ ion in perovsite iridates\cite{MorettiSala2014d}. The low-energy electronic structure is thus expected to be extremely sensitive to structural distortions acting on the local Ru$^{4+}$ crystalline environment.

Dramatic changes in the insulating ground state, including the appearance of superconductivity, have been indeed achieved by means of epitaxial strain \cite{Nobukane2017}, application of hydrostatic pressure to bulk crystals \cite{Alireza2010,Steffens2005} or internal chemical pressure \cite{friedt_structural_2001,fukazawa_filling_2001,Pincini2018a,Ricco2018}. The latter has been mainly realized by substitution of Ca with Sr \cite{friedt_structural_2001}, La \cite{fukazawa_filling_2001, Pincini2018a} or Pr\cite{Ricco2018}. This was found to suppress the MIT and drive the system into a metallic state. In particular, our recent neutron scattering measurements \cite{Pincini2018a} revealed that the different radii of the La$^{3+}$ ($r=1.22$~\AA) and Ca$^{2+}$ ($r=1.18$~\AA) ions\cite{Shannon1976} cause the compressed RuO$_6$ octahedra of pure \Ca{} to be progressively stretched along the $\mathbf{c}$ axis for increasing doping levels. This is expected to significantly change the local physics of the Ru$^{4+}$ ion and thus have a sizeable impact on the parent compound electronic structure. However, experimental studies on La-doped \Ca{} reported to date \cite{fukazawa_filling_2001,cao_ferromagnetic_2001,cao_ground-state_2000,Pincini2018a} have not addressed in detail the impact of the structural changes on the low-energy electronic structure.

\begin{figure}
	\centering
	\includegraphics[width=0.8\linewidth]{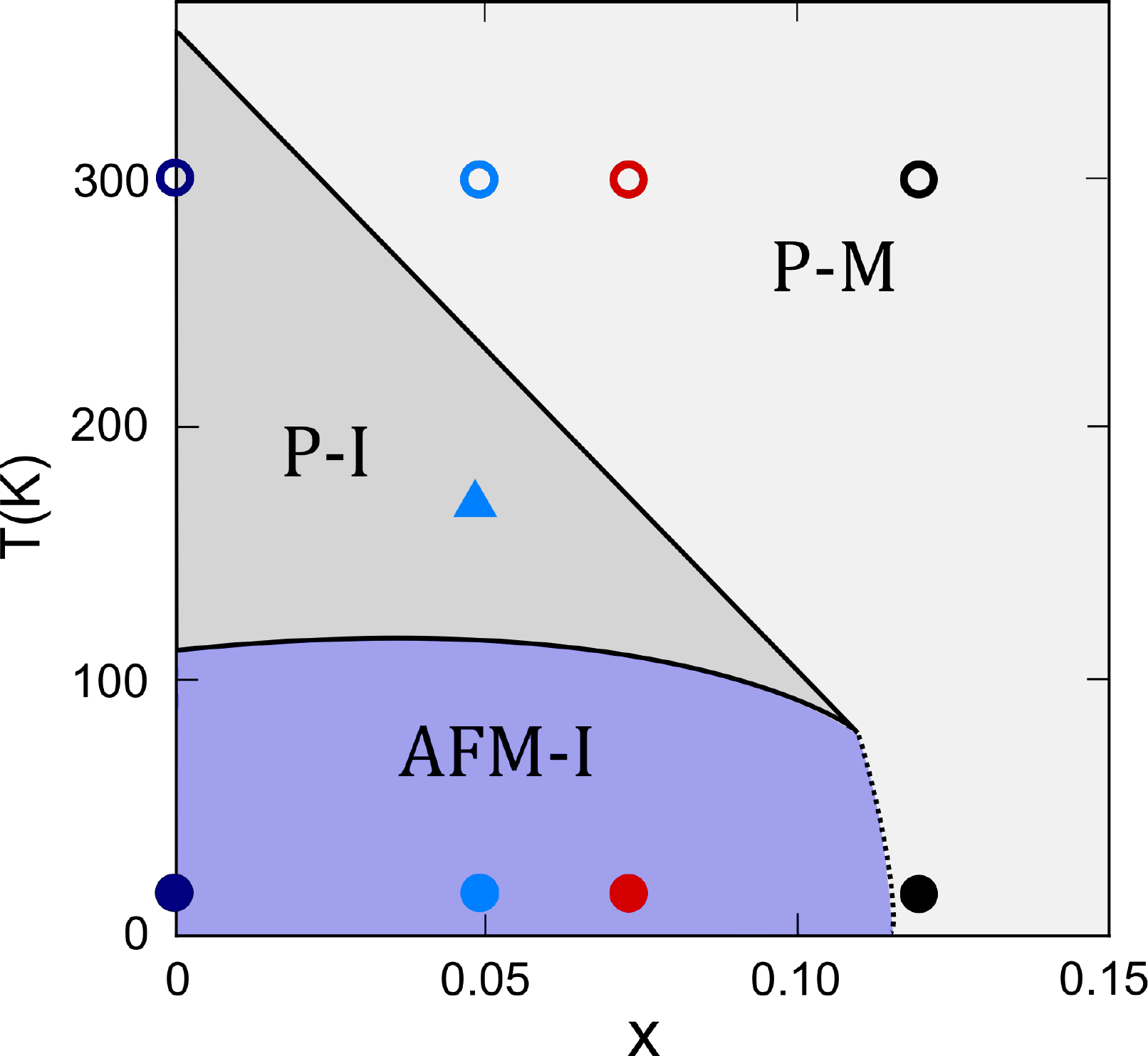}
	\caption{(Color online) \Cadoped{} temperature-doping phase diagram showing the paramagnetic metallic (P-M), paramagnetic insulating (P-I) and antiferromagnetic  insulating (AFM-I) phases. The filled and open symbols refer to the temperature and La content values in the insulating and metallic region, respectively, at which the XANES data were collected.}
	\label{phase_diagram}
\end{figure}

In this paper we report on an O $K$-edge X-ray absorption near edge structure (XANES) investigation of the Ru$^{4+}$ $t_{2g}$ orbital occupancy in \Cadoped{} in both the insulating and metallic regions of the phase diagram (Fig.~\ref{phase_diagram}) following a similar approach to the one already used in pure \Ca{}\cite{mizokawa_spin-orbit_2001,fatuzzo_spin-orbit-induced_2015} and Ba$_2$IrO$_4$\cite{sala_orbital_2014}. The XANES measurements are complemented by theoretical calculations based on a model Hamiltonian for the Ru-O-Ru cluster which includes electron-electron correlations, crystal field and SOC. In the insulating phase, the hole population of the $xy$ orbitals extracted from the absorption spectra is found to significantly increase as a function of the La concentration. Similar to the case of the parent compound\cite{mizokawa_spin-orbit_2001}, the hole occupancy of the doped samples with insulating ground states also appears to be temperature dependent, with an enhanced $xy$ hole population in the paramagnetic phase compared to the low-temperature antiferromagnetic (AFM) one. On the other hand, both temperature and doping are found to have very little impact in the metallic region. The results of our cluster calculations attribute the changes in the orbital population to the tetragonal crystal field tuning achieved by either La doping or temperature. In particular, the evolution from compressed to elongated RuO$_6$ octahedra as a function of increasing La content or temperature \cite{Pincini2018a} is found to cause a transfer of holes from the $yz,zx$ to the $xy$ orbitals, consistent with the XANES measurements.

\section{Experiment}\label{sec:experiment}

\subsection{Samples and methods}

Single crystals of \Cadoped{}, with $x=0$, $0.05(1)$, $0.07(1)$ and $0.12(1)$ [corresponding to the nominal dopings $x=0$, $0.05$, $0.10$ and $0.15$, respectively], were grown through the floating zone technique as described in Refs.~\onlinecite{Pincini2018a,Ricco2018}. The doping level was determined by means of energy-dispersive X-ray (EDX) spectroscopy and the bulk properties were characterized through magnetization and resistivity measurements \cite{Pincini2018a}. The structural properties were also investigated by means of single crystal neutron diffraction (see Ref.~\onlinecite{Pincini2018a} for further details). The parent compound ($x=0$) MIT is followed upon cooling below $T_N\approx 110$~K by a phase transition to a basal plane canted AFM state\cite{Alexander1999,Braden1998a,nakatsuji_ca_1997,cao_ground-state_2000,cao_ferromagnetic_2001,fukazawa_filling_2001,Fukazawa2000, Pincini2018a}. La substitution causes the MIT and N\'{e}el temperatures to decrease and be completely suppressed at a doping concentration slightly greater than $x=0.10$\cite{fukazawa_filling_2001, Pincini2018a}. This is schematically shown in the phase diagram of Fig.~\ref{phase_diagram}.

The XANES measurements were performed at the absorption branch of beamline I10 at Diamond Light Source (Didcot, UK). Absorption spectra were collected while scanning the incident, circularly-polarized, X-rays ($20 \times 100\,\mu m^2$ spot size) across the O $K$-edge energy ($543.1$~eV) for different values of the angle $\theta$ between the incident beam and the sample surface normal in the range $0-70^{^\circ}$. The degree of circular polarization provided by the \mbox{APPLE II} undulator was always larger than $99\%$\cite{Wang2012}. For each doping level, an equivalent data set was measured at both low ($\textrm{T}=10$~K) and room temperature, as indicated by the symbols in Fig.~\ref{phase_diagram}. The crystals were mounted on an electrically-grounded copper holder and inserted in the UHV sample environment with their crystallographic $\mathbf{c}$ axis aligned parallel to the incident beam for $\theta=0^{\circ}$. The absorption was simultaneously measured in both total-electron yield (TEY) and total fluorescence yield (TFY) detection mode. The analysis of both sets of data led to similar conclusions and only the TEY measurements are shown in the present work. Cleaving the crystals \textit{in-situ} did not result in any appreciable difference in the absorption spectra with respect to the non-cleaved ones, thus excluding any impact of surface contamination.


\subsection{Results}

\subsubsection{Insulating phase}

\begin{figure}
	\centering
	\includegraphics[width=\linewidth]{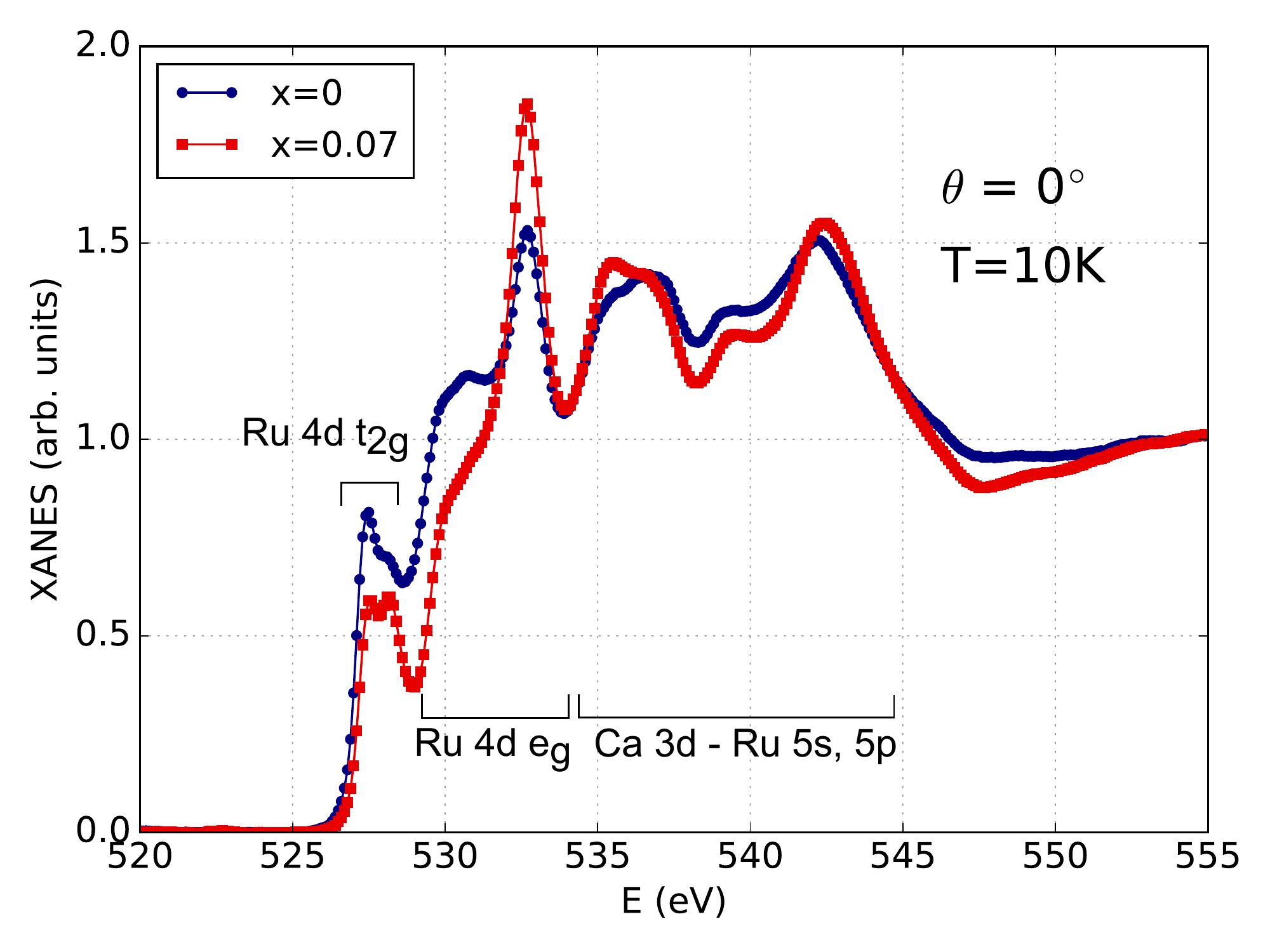}
	\caption{(Color online) O $K$-edge XANES spectra for the parent (blue circles) and $x=0.07$ (red squares) compound collected at normal incidence and low temperature. The spectra were normalized using the spectral weight at $E>553$~eV.}
	\label{long_spectra}
\end{figure}

\begin{figure}
	\centering
	\includegraphics[width=0.8\linewidth]{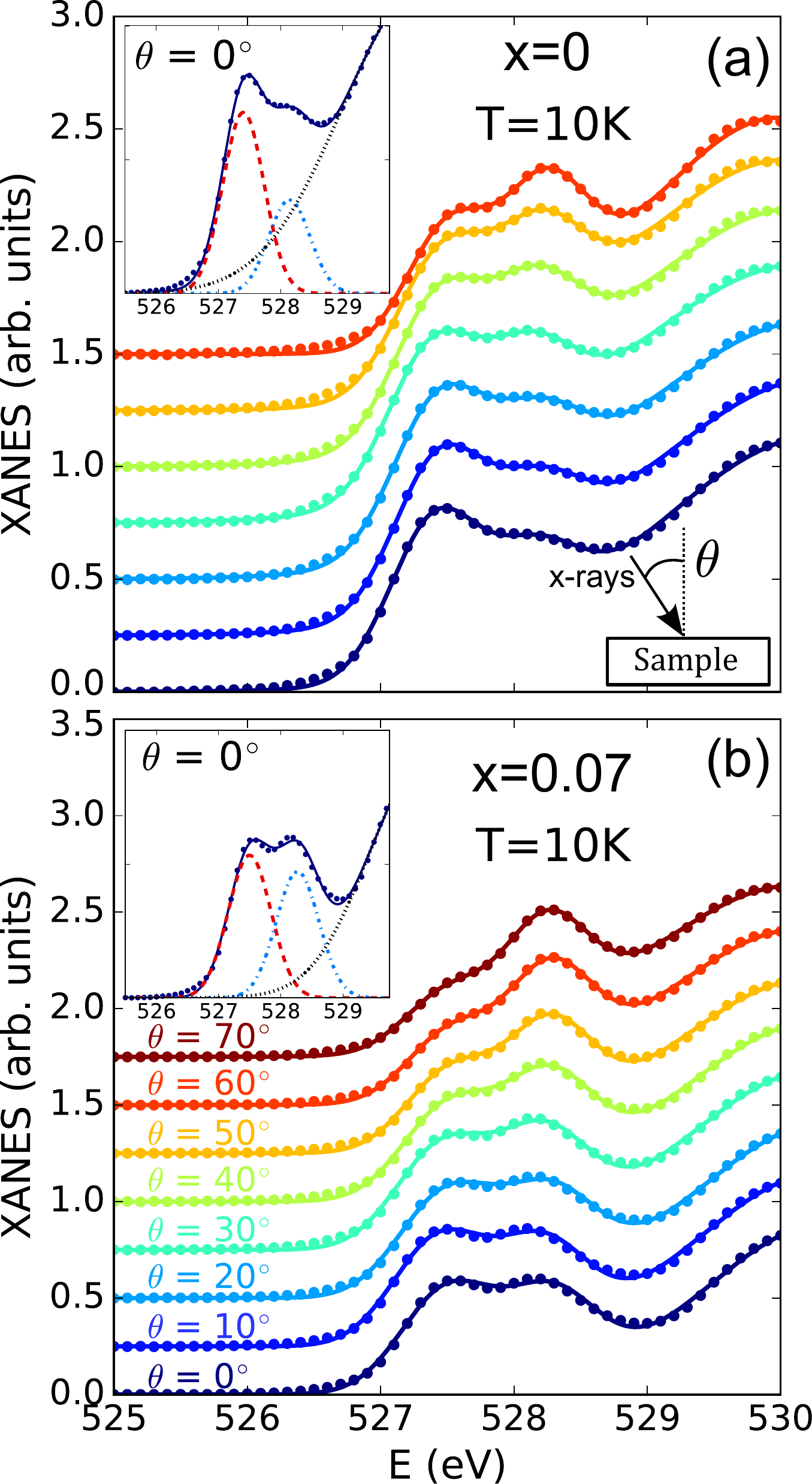}
	\caption{(Color online) XANES spectra collected at different values of the angle $\theta$ between the incident X-ray beam and the sample surface normal in the (a) parent and (b) $x=0.07$ sample at low temperature. The insets in (a) and (b) show the fit detail of the spectra measured at normal incidence ($\theta=0^{\circ}$). The measured absorption (filled circles) was fitted to the sum (solid line) of three Gaussian profiles modelling the apical (red dashed line) and in-plane (blue dot-dash line) O $2p$ - Ru $t_{2g}$ features and the $e_g$ region of the spectrum (black dotted line). The corresponding angular dependence of the ratio of in-plane to apical intensities is reported in Fig.~\ref{angular_dependence}. Spectra at different angles were normalized using the spectral weight at $E>553$~eV.}
	\label{spectra_vs_angle_vertical}
\end{figure}

Low-temperature O $K$-edge spectra at normal incidence for parent \Ca{} and the $x=0.07$ sample are shown in Fig.~\ref{long_spectra}. As first reported by Mizokawa et al\cite{mizokawa_spin-orbit_2001}, the XANES signal shows several features arising from the hybridization of the O $2p$ orbitals with the Ru $4d$ $t_{2g}$ ($E<530$~eV), Ru $4d$ $e_g$ ($530\;\textrm{eV}<E<535$~eV) and Ca $3d$ / Ru $5s,\,5p$ ($E>535$~eV) orbitals. La doping has a significant impact on the relative intensities of the different features, thus revealing the occurrence of sizeable changes in the empty density of states of the Ru$^{4+}$ ion. In the present work, we focus our attention on the $t_{2g}$ region of the spectrum, which consists of two separate features centred at around $E\approx 527.5$~eV and $E\approx528.2$~eV. 
Following the peak assignments in pure \Ca{}\cite{mizokawa_spin-orbit_2001,fatuzzo_spin-orbit-induced_2015,Das2018} and Sr$_2$RuO$_4$\cite{Schmidt1996,Moon2006} and the band structure calculations in \Ca{}\cite{Fang2004a} and Sr$_2$RuO$_4$\cite{Singh1995}, we attribute the two features to the hybridization of the Ru $t_{2g}$ orbitals with the apical and in-plane O $2p$ orbitals of the RuO$_6$ octahedra, respectively. Their relative intensity depends on the angle of incidence of the X-ray beam: this is clearly shown in Fig.~\ref{spectra_vs_angle_vertical}, where spectra collected at low temperature for different $\theta$ values in the range $0-70^{\circ}$ are reported for the parent and $x=0.07$ crystal. The in-plane/apical intensity ratio tends to increase as $\theta$ is increased away from normal incidence. A similar effect also occurs upon doping, as can be seen by comparing spectra collected at the same $\theta$ value in the parent and doped sample.

\begin{figure}
	\centering
	\includegraphics[width=\linewidth]{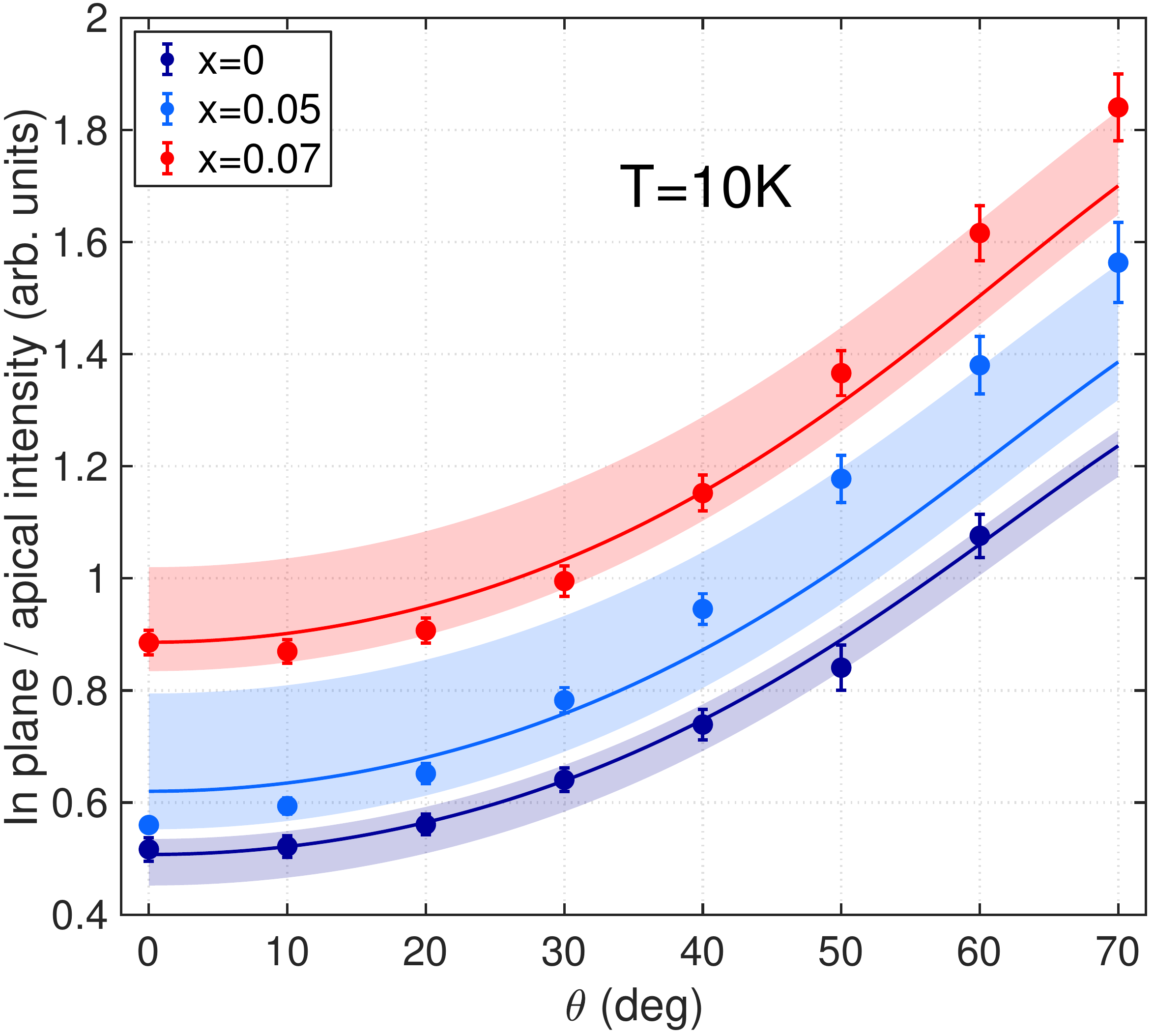}
	\caption{(Color online) Ratio of the intensities of the in-plane to apical O $2p$ - Ru $t_{2g}$ features of the XANES spectra in the parent (dark blue circles), $x=0.05$ (light blue triangles) and $x=0.07$ (red squares) compounds at low temperature. The filled symbols were extracted through a fit of the measured XANES spectra analogous to the ones shown in Fig.~\ref{spectra_vs_angle_vertical} (see Supplemental Material\cite{supplemental_material} for the XANES spectra of the $x=0.05$ sample), while the solid lines correspond to the best fit to the cross sections of Eq.~(\ref{ratio_equ}), and the shaded areas to estimated confidence intervals in these fits. The resulting $n_{xy}/n_{yz,zx}$ ratios are reported in Fig.~\ref{hole_ratio}.}
	\label{angular_dependence}
\end{figure}

A quantitative analysis of the angular dependence was achieved by fitting the low-energy region of the spectra ($E<530$~eV) to the sum of three Gaussian peaks modelling the apical and in-plane O $2p$ - Ru $t_{2g}$ features and the $e_g$ region of the spectrum at higher energy (see insets in Fig.~\ref{spectra_vs_angle_vertical}). The results of the fits for the different insulating samples at low temperature are reported in Fig.~\ref{angular_dependence}. A clear evolution of the angular dependence is seen as a function of doping, with the in-plane/apical intensity ratio increasing with the La content. 

\begin{figure}
	\centering
	\includegraphics[width=0.95\linewidth]{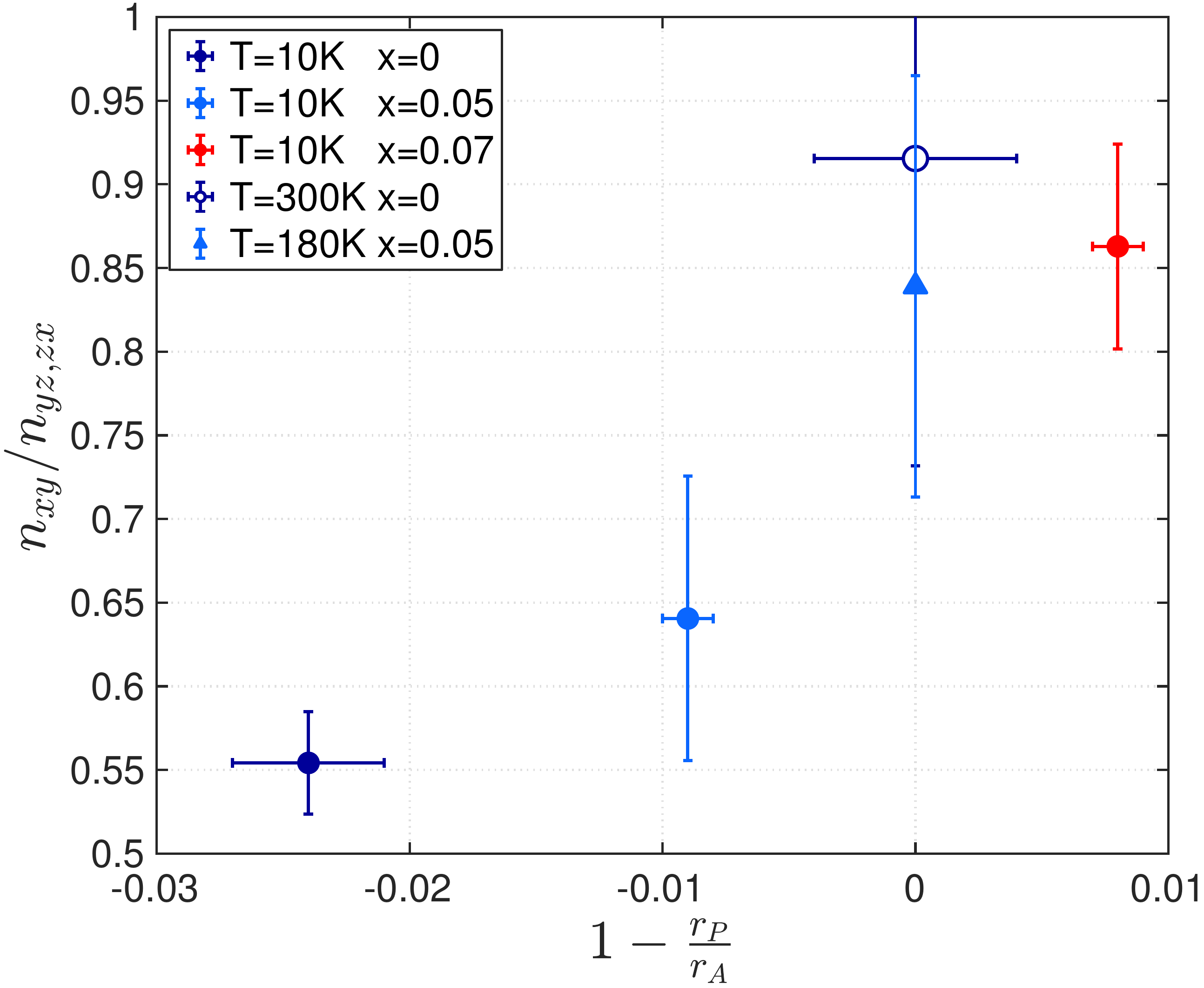}
	\caption{(Color online) Hole occupancy ratio $n_{xy}/n_{yz,zx}$ at different temperatures as a function of tetragonality $1-\left(r_P / r_A\right)$, extracted from the angular dependence of the ratio of the in-plane to apical O $2p$ - Ru $t_{2g}$ intensities in the XANES spectra in the insulating region of the phase diagram (see Fig.~\ref{angular_dependence} for the $\textrm{T}=10$~K data set and the Supplemental Material\cite{supplemental_material} for the measurements at higher temperature). The bond lengths $r_{A,P}$ were obtained from Ref.~\onlinecite{Pincini2018a} for all doping and temperature combinations except $x=0.05$, $\textrm{T}=180$~K, for which it was assumed $r_A=r_P$. The symbols follow the same convention used in Fig.~\ref{phase_diagram}. The vertical error bars correspond to the estimated confidence intervals shown in Fig.~\ref{angular_dependence}, while the horizontal error bars correspond to the uncertainties in the bond lengths\cite{Pincini2018a}.}
	\label{hole_ratio}
\end{figure}

The dependence of the intensity ratio on the incident angle is determined by the dipole matrix elements of the O $1s\rightarrow 2p$ transition. Following the minimal hybridization model already exploited for \Ca{}\cite{mizokawa_spin-orbit_2001,fatuzzo_spin-orbit-induced_2015}, the $\theta$ dependence of transitions to the O $2p_{x}$, $2p_{y}$ and $2p_z$ orbitals  for circularly polarized light\footnote{Here, we consider the representative case where the crystallographic $[0\,1\,0]$ direction lies along the $\theta$ rotation axis for the expression of the angular dependence of the transitions to the $2p$ orbitals. The resulting angular dependence of the intensity of the XANES features does not depend on this assumption.} is given by $\frac{1}{2}\cos^2{\theta}$, $\frac{1}{2}$ and $\frac{1}{2}\sin^2{\theta}$, respectively\cite{mizokawa_spin-orbit_2001,sala_orbital_2014}. The $2p_{x,y}$ and $2p_z$ orbitals of the in-plane O atoms hybridize with the Ru $xy$ and $yz,zx$ orbitals, respectively. On the other hand, the $2p_x$ ($2p_y$) orbital of the apical O atoms hybridizes with the Ru $zx$ ($yz$) orbital. The XANES intensities of the apical ($I_A$) and in-plane ($I_P$) oxygen features are thus given by the following relations\cite{mizokawa_spin-orbit_2001}
\begin{equation}
\begin{array}{cc}
\displaystyle I_A(\theta)\propto r_A^{-3.5}\frac{1}{2}(\cos^2{\theta}+1)n_{yz,zx}\\\\
\displaystyle I_P(\theta)\propto r_P^{-3.5}\left[\frac{1}{2}(\cos^2{\theta}+1)n_{xy}+\frac{1}{2}\sin^2{\theta}\,n_{yz,zx}\right]
\end{array}
\label{apical_inplane_equ}
\end{equation}
giving a ratio
\begin{equation}
\begin{array}{cc}
\displaystyle \frac{I_P}{I_A}\propto \left(\frac{r_A}{r_P}\right)^{3.5}\left[\frac{n_{xy}}{n_{yz,zx}}+\frac{\sin^2{\theta}}{\cos^2{\theta}+1}\right]
\end{array}
\label{ratio_equ}
\end{equation}
where $r_A$ ($r_P$) is the apical (in-plane) Ru-O bond length and $n_{xy}$ ($n_{yz,zx}=n_{yz}+n_{zx}$) is the number of holes in the Ru $xy$ ($yz+zx$) orbitals. In Eq.~(\ref{apical_inplane_equ}), we assumed that the hybridization strength decays with the Ru-O bond length as $r^{-3.5}$, similar to a previous XANES investigation on the parent compound\cite{fatuzzo_spin-orbit-induced_2015}. In addition, the cross sections of Eq.~(\ref{apical_inplane_equ}) do not depend on the particular in-plane orientation of the crystallographic $\mathbf{a}$ and $\mathbf{b}$ axes, which was not defined in our measurements.

It should be noted that determining the Ru $4d$ orbital occupancy through the direct measurement of the oxygen $2p$ orbitals is not a priori obvious since one needs to understand the correspondence between the hole distribution in the two orbitals. However, as will be discussed later, our theoretical calculations indicate that the hole occupancy of the O $2p$ and Ru $4d$ orbitals show the same general trend, which validates the minimal hybridization model used to analyze the data.

Equation~(\ref{ratio_equ}) was used to fit the measured angular dependences and extract the hole occupancy ratio $n_{xy}/n_{yz,zx}$ (note that the arbitrary scale factor which relates the cross sections to the intensity derived from the XANES spectra means that only the ratio is accessible through a fit of the data rather than the separate quantities $n_{xy}$ and $n_{yz,zx}$). The best-fit curves for the insulating samples at low temperature are plotted as solid lines in Fig.~\ref{angular_dependence}, with estimated confidence intervals in the fits represented by shaded areas. These fits were obtained considering $r_A=1.971,\,1.987,\,2.007\,\textrm{\AA}$ and $r_P=2.017,\,2.005,\,1.990\,\textrm{\AA}$ for $x=0,\,0.05\,,0.07$, respectively\cite{Pincini2018a}. An angular offset was also included as a free fitting parameter to account for small misalignments of the sample surface normal with respect to the incident X-ray direction. Equation~(\ref{ratio_equ}) provides a good description of the trends in the experimental data: the resulting $n_{xy}/n_{yz,zx}$ values are reported in Fig.~\ref{hole_ratio} as a function of the tetragonality $1-\left(r_P / r_A\right)$. Here, the values extracted from the XANES spectra measured in the parent compound at room temperature and the $x=0.05$ sample in the paramagnetic insulating (P-I) phase ($\textrm{T}=180$~K) are also shown. Given the lack of detailed structural information at $\textrm{T}=180$~K, we set 
$r_A=r_P$ at this temperature.


\begin{figure}
	\centering
	\includegraphics[width=\linewidth]{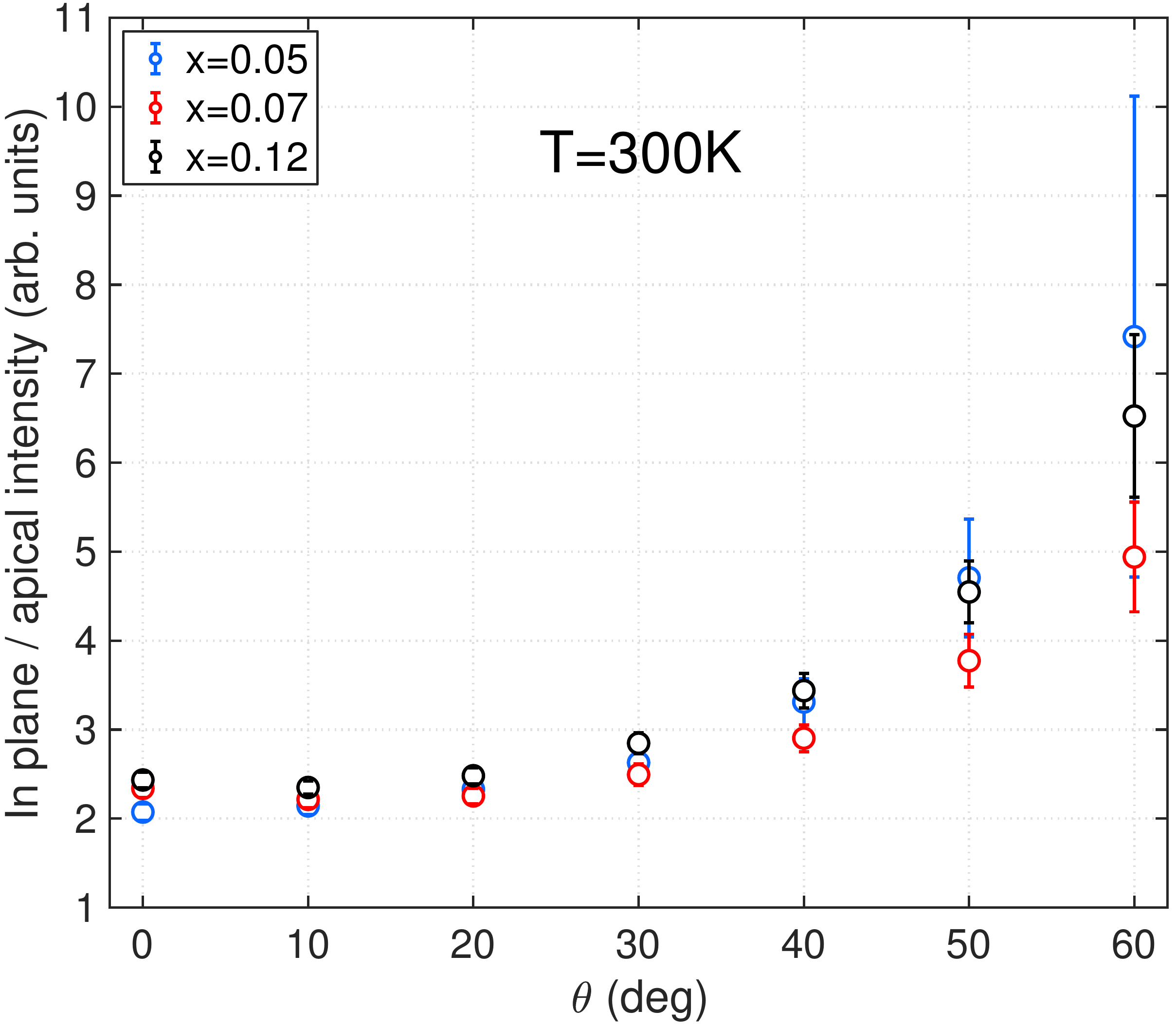}
	\caption{{(Color online) Ratio of the intensities of the in-plane to apical O $2p$ - Ru $t_{2g}$ features of the XANES spectra in the $x=0.05$ (dark blue circles), $x=0.07$ (light blue triangles) and $x=0.12$ (red squares) samples at room temperature. The filled symbols were extracted through the fit of the measured XANES spectra analogous to Fig.~\ref{angular_dependence} (see the Supplemental Material\cite{supplemental_material} for the corresponding spectra).}}
	\label{angular_dependence_metallic}
\end{figure}

Our measurements indicate that, at low temperature, the $n_{xy}/n_{yz,zx}$ value increases from $0.55(3)$ in the parent compound up to $0.86(6)$ in the $x=0.07$ sample. Consistent with an early report on pure \Ca{}\cite{mizokawa_spin-orbit_2001}, the hole occupancy of the $xy$ orbitals is also found to increase up to $n_{xy}/n_{yz,zx}\approx1$ upon warming to the P-I phase in the $x=0$ ($\textrm{T}=300$~K) and $x=0.05$ ($\textrm{T}=180$~K) samples. Previous XANES studies on the parent compound reported somewhat smaller $n_{xy}/n_{yz,zx}$ values in the AFM phase\cite{mizokawa_spin-orbit_2001,fatuzzo_spin-orbit-induced_2015}. In particular, Mizokawa et al\cite{mizokawa_spin-orbit_2001} found $n_{xy}/n_{yz,zx}\approx0.3$ using circularly polarized light at $\textrm{T}=90$~K, while XANES measurements performed with linearly polarized X-rays in normal and grazing incidence geometry at $\textrm{T}=20$~K reported a value $0.15< n_{xy}/n_{yz,zx}< 0.2$\cite{fatuzzo_spin-orbit-induced_2015}. A significant change in the orbital population between $90$ and $20$~K seems unlikely given the substantial insensitivity of the \Ca{} crystal structure on the temperature below the N\'{e}el transition\cite{Braden1998a}. The variability between different studies might stem from small variations in the oxygen content: these are likely to affect the orbital population at the Ru site due to the covalent character of the Ru-O bond \cite{Braden1998a}. Although the origin of the discrepancy remains an open issue, this does not affect the validity of our conclusions.


\subsubsection{Metallic phase}

The in-plane and apical features of the XANES spectra exhibit a markedly different behaviour in the metallic phase (see Fig.~\ref{phase_diagram}). In particular, the apical feature is found to be less pronounced than the in-plane one at all $\theta$ values. Figure~\ref{angular_dependence_metallic} shows the angular dependence measured in the the doped crystals at room temperature (see the Supplemental Material\cite{supplemental_material} for the corresponding XANES spectra). In contrast to the insulating phase, the doping concentration does not significantly affect the angular dependence of the XANES features. The same is true for temperature, as shown in the low-temperature data collected in the $x=0.12$ sample\cite{supplemental_material}. Equation~(\ref{ratio_equ}) does not provide a satisfactory fit of the data in this case. This is not surprising, since the simple hybridization model is based on an atomic orbitals picture and cannot properly account for the itinerant character of the Ru $4d$ electrons in the metallic state.

\section{Theoretical calculations}\label{sec:calculations}

\begin{figure}
	\includegraphics[width=0.95\columnwidth]{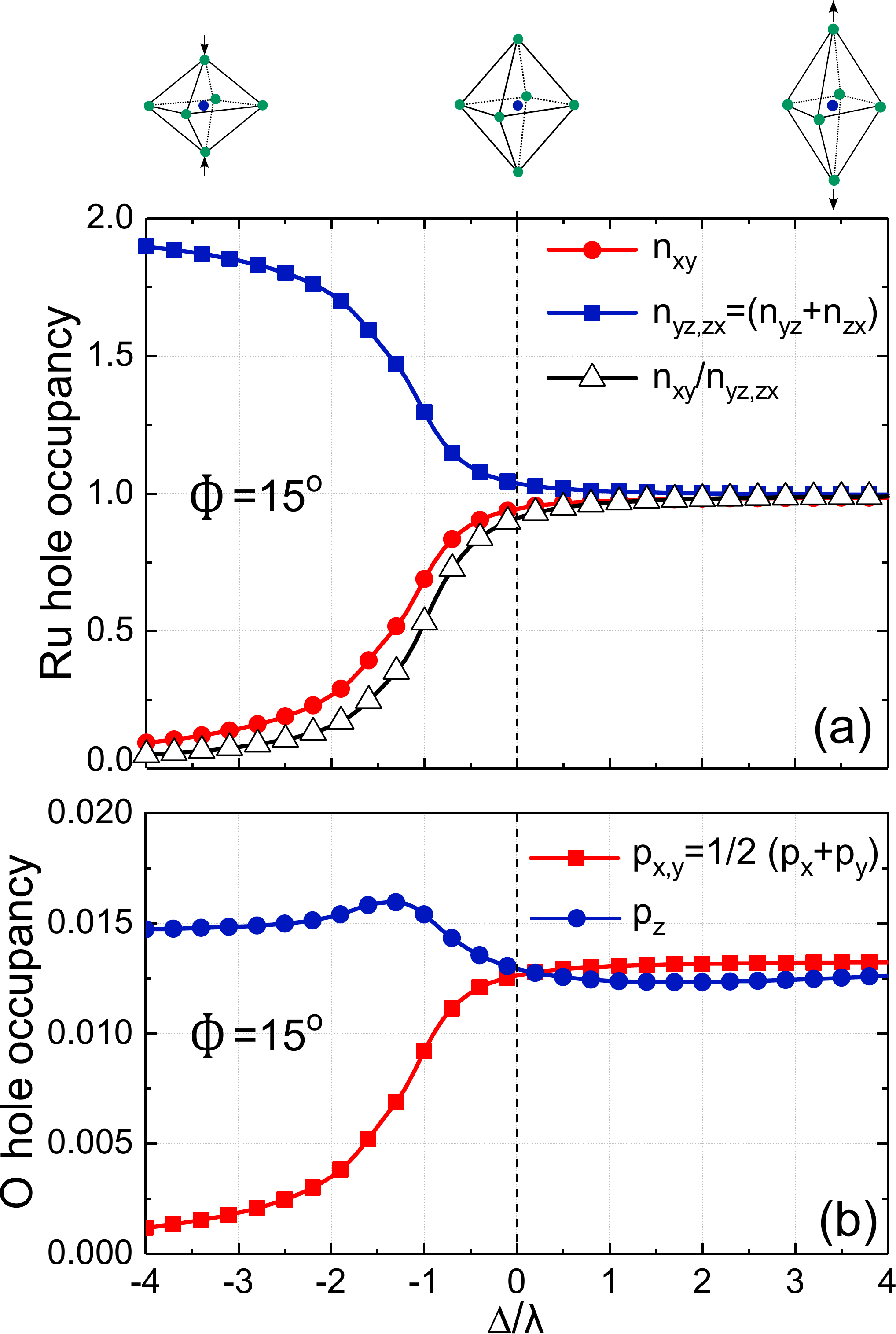}
	\caption{(Color  online) Calculated hole occupancy of the (a) Ru $t_{2g}$ and (b) O $2p$ orbitals as a function of the ratio $\Delta/\lambda$ between the tetragonal crystal field potential and the SOC. The calculations were performed considering the ground state of the Ru-O-Ru cluster with $J_H=0.4$~eV, $U=2.2$~eV and $\phi=15^\circ$, where $\phi$ is the RuO$_6$ octahedra in-plane rotation. The diagram at the top of the figure schematically illustrates the tetragonal distortion of the RuO$_6$ octahedra for the different values of $\Delta/\lambda$.}
	\label{THEORY}
\end{figure}

The recent report of tetragonal field tuning achieved by La substitution \cite{Pincini2018a} points towards a structural origin for the changes in the hole occupancy extracted from the XANES data. Following this reasoning, we investigated the evolution of the orbital-dependent hole occupancy of the Ru $4d$ and O $2p$ electronic states as a function of the ratio $\Delta/\lambda$ of the tetragonal crystal field ($\Delta$) and SOC constant ($\lambda$). \Ca \, is a spin-orbit coupled Mott insulator with strong sensitivity to structural changes \cite{Nobukane2017,Alireza2010,Steffens2005,friedt_structural_2001,fukazawa_filling_2001,Pincini2018a,Ricco2018}. Therefore, it is of primary importance to include into the microscopic modelling the atomic Coulomb interaction, the SOC, the $p$-$d$ charge transfer processes and the octahedral distortions. Moreover, since the XANES measurements of Sec.~\ref{sec:experiment} give direct insight into the oxygen $2p$ states, one relevant issue here is to asses how the Ru $4d$ orbital occupation is related to O $2p$ one. Since both the SOC and the octahedral rotations mix the orbital degrees of freedom, it is not a priori obvious to deduce how the changes in the hole distribution in the $4d$ orbitals are related to the ones in the $2p$ states at the oxygen site. 

The calculations were performed by solving a model Hamiltonian for the Ru-O-Ru cluster that is able to capture the main electronic processes impacting the $4d$ and $2p$ orbital occupations in the AFM insulating phase of \Cadoped. An analogous approach was recently successfully used to interpret the spectrum of Ru$^{4+}$ electronic excitations probed by means of RIXS at the O $K$ edge \cite{Das2018}. The model Hamiltonian for the relevant bands close to the Fermi level for the electrons within the ruthenium-oxygen plane is based on the interaction terms at the Ru and O sites and the charge transfer processes for the Ru-O electronic connectivity. The local ruthenium Hamiltonian $H_{\mathrm{loc}}$\cite{Cuoco2006a,Cuoco2006} includes
the complete Coulomb interaction for the t$_{2g}$ electrons, the SOC and the tetragonal crystal field potential. The various terms are generally expressed by:

\begin{eqnarray}
H_{\text{el-el}}(i) 
&=&
U\sum n_{i\alpha \uparrow }n_{i\alpha \downarrow}
-2J_{\mathrm{H}}\sum\limits_{\alpha <\beta }\bf{S}_{i\alpha }\cdot 
\bf{S}_{i\beta }
+\nonumber \\&&+
\left(U-\frac{5 J_{\mathrm{H}}}{2}\right)\sum\limits_{\alpha <\beta }n_{i\alpha }n_{i\beta }
+\nonumber \\&&+J_{H}\sum\limits_{\alpha <\beta }d_{i\alpha \uparrow }^{\dagger }d_{i\alpha
	\downarrow }^{\dagger }d_{i\beta \uparrow }d_{i\beta \downarrow }, \nonumber \\
H_{\mathrm{SOC}}(i) &=&\lambda \sum\limits_{\alpha ,\sigma }\sum_{\beta ,\sigma
	^{^{\prime }}} d_{i\alpha \sigma }^{\dagger } (\bf{l}_{\alpha\beta }\cdot \bf{s}_{\sigma \sigma ^{^{\prime }}})
d_{i\beta
	\sigma ^{^{\prime }}} ,\nonumber \\
H_{\mathrm{cf}}(i) &=&\varepsilon_{xy} n_{i,xy}+\varepsilon_{z}\left(
n_{i,xz}+n_{i,yz}\right), \nonumber \\
H_{\mathrm{loc}}(i) &=&H_{\text{el-el}}(i)+H_{\mathrm{SOC}}(i)+H_{\mathrm{cf}}(i)\,
\end{eqnarray}

\noindent where $i$ labels the site and $\alpha,\beta $ are indices running over the three orbitals in the t$_{2g}$
sector, i.e. $\alpha,\beta \in \{d_{xy},d_{xz},d_{yz}\}$, and $d_{i\alpha \sigma }^{\dagger }$ is
the creation operator of an electron with spin $\sigma $ at the site $i$ in
the orbital $\alpha$. The interaction is parametrized by the intra-orbital Coulomb interaction $U$ and the Hund's coupling  $J_{\mathrm{H}}$. The strength of the tetragonal distortions is expressed by the amplitude $\Delta$, with $\Delta=(\varepsilon_{xy}-\varepsilon_z)$. A negative (positive) $\Delta$ corresponds to a flat (elongated) octahedral configuration with a tendency to different orbital occupations in the two regimes.

Furthermore, we consider the ruthenium-oxygen hopping, which includes all the symmetry-allowed terms according to the Slater-Koster rules \cite{Harrison2012,Brzezicki2015} for a given bond connecting a ruthenium to an oxygen atom along a given symmetry direction. Here, we allow for a rotation of the octahedra around the $\mathbf{c}$ axis assuming that the planar Ru-O-Ru bond can form a generic angle $\beta=(180^{\circ}-\phi)$. The case with $\phi=0$ corresponds to the tetragonal undistorted bond, while a non-vanishing value of $\phi$ arises when the RuO$_6$ octahedra are rotated around the $\mathbf{c}$ axis. A value of $\phi=15^{\circ}$ was considered in the present analysis, which is in the range of the experimentally observed octahedral rotation for \Ca \cite{Pincini2018a}.

Concerning the employed electronic parameters, we took as a reference those previously used for \Ca. In the latter, the octahedra become flat below the structural transition \cite{Braden1998a,Pincini2018a}: $\Delta$ is thus negative and, according to first-principle calculations or estimates employed to reproduce the RIXS \cite{Das2018} or inelastic neutron scattering\cite{jain_higgs_2017} spectra, its magnitude is $\sim200-300$~meV. The material specific values $\lambda=0.075$ eV, $U=2$ eV and $J_{H}$ in the range $0.35-0.5$~eV \cite{mizokawa_spin-orbit_2001, Veenstra2014} were also considered. Similar values for $\Delta$, $U$ and $J_H$ have been used for band structure calculations in \Ca \cite{Sutter2017a}, while a ratio $|\Delta|/ \lambda$  in the range $\sim3-4$ was found to correctly reproduce the spin excitations observed by inelastic neutron scattering \cite{jain_higgs_2017}. For the hopping amplitudes, we considered a representative set of electronic parameters for the Ru-O-Ru cluster that is consistent with typical amplitudes obtained from first-principle calculations on ruthenium oxides\cite{Fang2004a,Gorelov2010b,Das2018,Malvestuto2013,Granata2016,Forte2010}. Given the limited impact of La substitution on the Ru-O-Ru bond angle and octahedral tilt away from the $\mathbf{c}$ axis \cite{Pincini2018a}, we assumed that the hopping amplitudes are not quantitatively affected by the La concentration.

The evolution of the Ru $xy$, $yz$ and $zx$ orbital population can be conveniently described in terms of the expectation values ${n}_{\alpha}$ of the hole occupancy operators $\tilde{n}_{\alpha}$ ($\alpha=xy,yz,zx$). These are reported in Fig.~\ref{THEORY}(a) as a function of $\Delta/\lambda$. Here, a value $n_{xy}/n_{yz,zx}=0.5$ for the ratio of the $xy$ and $yz+zx$ hole occupancies corresponds to an equal population of holes in each of the $xy$, $yz$ and $zx$ states. As expected, a transfer of holes from the $yz,zx$ to the $xy$ orbitals takes place in going from flat ($\Delta<0$) to elongated ($\Delta>0$) RuO$_6$ octahedra. In particular, we observe that the crossover from $n_{xy}/n_{yz,zx}<0.5$ to $n_{xy}/n_{yz,zx}>0.5$ starts already in the regime of flat octahedra ($\Delta/\lambda\sim-1$), after which the system evolves rapidly to a configuration with an equal number of holes in the $xy$ and $yz+zx$ orbitals. The latter corresponds to $n_{xy}/n_{yz,zx}=1$ and is almost completely realized already for $\Delta=0$. We find that such behavior is quite robust and weakly depends on the amplitude of the Coulomb interaction. More specifically, a variation of the local atomic correlations leads to a shift of the onset of the hole transfer from the $xy$ to the $yz,zx$ orbitals in the direction of octahedral compression. 

Another interesting aspect emerging from our calculations concerns the correspondence between the hole distribution in the $2p$ orbitals at the oxygen site and the one of the Ru $4d$ states. Our results show that the number of holes in the planar $p_{x,y}$ ($p_z$) orbitals exhibits the same trend of the $xy$ ($yz,zx$) states, as a function of the tetragonal distortions (Fig.~\ref{THEORY}(b)). This is particularly relevant as it justifies the possibility of extracting the evolution of the hole distribution in the Ru bands from the spectroscopic analysis performed at the O sites by means of Eq.~(\ref{ratio_equ}).

\section{Discussion}

The results of the cluster calculations presented in Sec.~\ref{sec:calculations} clearly identify the tetragonal crystal field tuning induced by La substitution (or temperature) that we have recently reported\cite{Pincini2018a} as the main cause of the orbital hole occupancy evolution extracted from the XANES measurements in the insulating phase. In particular, the neutron scattering investigation of Ref.~\onlinecite{Pincini2018a} revealed that the internal chemical pressure caused by La doping results in a progressive elongation of the RuO$_6$ octahedra along the apical Ru-O direction. The latter are compressed by about $-2.4\%$ in the parent compound \cite{Braden1998a,Pincini2018a}. The compression reduces to $-0.9\%$ in the $x=0.05$ sample, while the octahedra are elongated by about $0.8\%$ for $x=0.07$ \cite{Pincini2018a}. Temperature was found to have a similar effect in the parent compound insulating phase: the tetragonal distortion is released upon warming, leading to almost regular octahedra at $\textrm{T}=300$~K\cite{Braden1998a,Pincini2018a}. 

Consistent with the calculations displayed in Fig.~\ref{THEORY}(a), the ratio $n_{xy}/n_{yz,zx}$ increases in going from the parent to the doped compounds and is larger in the P-I phase than in the AFM one. In particular, the $xy$ hole population at low temperature is significantly enhanced in the case of the $x=0.07$ elongated octahedra with respect to the $x=0,\,0.05$ compressed ones. The attribution of the observed evolution to structural effects is further confirmed by the insensitivity of the XANES spectra of the metallic samples to both temperature and La content. Our recent neutron diffraction study \cite{Pincini2018a} indeed found that the RuO$_6$ octahedra are elongated by about $4.3-4.5\%$ in the metallic region of the phase diagram regardless of the temperature or doping level.

The results presented in Fig.~\ref{THEORY} also contain important information with relevance to the impact of the spin-orbit coupling on the Ru $4d$ orbital occupancy. It explicitly demonstrates how the orbital population of the Ru $t_{2g}$ states are tuned by the tetragonal crystal field potential in the presence of SOC and Coulomb interactions, providing direct insight into the energy scales expressing the competition between these parameters. Moreover, the faster transfer of holes from the $zx$, $yz$ to the $xy$ orbitals already in the regime of flat octahedra ($\Delta / \lambda \sim -1$) is a direct consequence of SOC as it allows the mixing of these orbitals. In this case, SOC acts to weaken the effect of the crystal field potential. The emptying out of the $xy$ orbital, then, leads to a breakdown of the Mott insulating state and the formation of a metallic state which is electronically similar to the high temperature phase of \Ca{}, where it has elongated octahedra.

Despite the structural modifications induced by La doping unveiled by neutron diffraction\cite{Pincini2018a}, the corresponding effect on the Ru$^{4+}$ electronic structure is not a priori obvious. Indeed, La substitution can also introduce free carriers at the Ru sites. In this case, one would observe a decrease in the hole occupation of $yz,zx$ states without significant modifications in the $xy$ hole density. Such behavior would thus result in an increase in the $n_{xy}/n_{yz,zx}$ ratio upon doping, similar to what is expected when considering structural effects alone (Fig.~\ref{THEORY}). However, recent investigations on La-doped\cite{Pincini2018a} and Pr-doped\cite{Ricco2018} \Ca\ strongly suggest that, in contrast to lightly doped cuprates\cite{Damascelli2003} and iridates\cite{delaTorre2015,Kim2014}, the doped electrons remain fully localized in the low-temperature insulating phase, confirming the predominance of the structural effects on the ground-state orbital occupancy of the Ru$^{4+}$ ion.

The ground-state properties of the Ru$^{4+}$ ion in \Ca\,were previously described in terms of a minimal Hamiltonian including tetragonal crystal field and SOC, which was used to account for the electronic excitation spectrum probed by O $K$-edge RIXS\cite{fatuzzo_spin-orbit-induced_2015}. The model of Ref.~\onlinecite{fatuzzo_spin-orbit-induced_2015} represents an oversimplification of the electronic properties of \Ca\, as, in contrast with the cluster calculations of Sec.~\ref{sec:calculations}, it neglects many-body effects of the system of four Ru$^{4+}$ $t_{2g}$ electrons and the interaction with the ligand oxygens. Nonetheless, this minimal Hamiltonian also predicts an enhancement of the $xy$ hole populations in going from octahedral compression to octahedral elongation\cite{supplemental_material}, in qualitative agreement with our results

Although the trends seen in our data qualitatively follow the predictions of the model (Sec.~\ref{sec:calculations}), the measured absolute values of $n_{xy}/n_{yz,zx}$  are only in partial agreement with the calculated ones. This is mainly due to the fact that the $n_{xy}/n_{yz,zx}$ values derived through the minimal hybridization model of Eq.~(\ref{ratio_equ}) are only approximate and do not allow a precise quantitative estimation. Despite the quantitative discrepancies, the trend confirmed by the calculations highlights the sensitivity of the spin-orbit entangled ground state of the Ru$^{4+}$ ion to structural distortions.

\section{Concluding remarks}

In conclusion, our XANES investigation has revealed that the elongation of the RuO$_6$ octahedra induced by either La substitution or temperature increase results in an enhancement of the $xy$ hole population of the Ru$^{4+}$ ground-state wave function in the insulating region of the temperature-doping phase diagram. On the other hand, the hole occupancy shows little variation with temperature and doping in the metallic phase, consistent with the lack of significant structural changes. The sensitivity of the orbital population of the Ru$^{4+}$ ground-state to the local crystalline environment has been shown to directly result from the peculiar entanglement of $xy$, $yz$ and $zx$ orbitals caused by the interplay of electronic correlations, crystal field and SOC of $4d$ electrons. Our findings confirm the subtle nature of the low-energy Hamiltonian of perovskite ruthenates, where, despite the presence of a rather weaker spin-orbit interaction compared to the case of iridium oxides, a peculiar coupling between orbital and lattice degrees of freedom still arises from the competition between tetragonal field and SOC. Moreover, the results highlight the unique impact of La-substitution on the Mott-band insulating state of {\Ca}, where the low-temperature metallic phase is stabilized by the structurally induced redistribution of holes in the $t_{2g}$ orbitals rather than by the injection of electrons.


\begin{acknowledgements}

The authors would like to thank S. Ricc\`{o} for her great help during the sample growth and characterization and J. Chang, S. Ricc\`{o}, F. Baumberger, M. M. Sala, M. Rossi and S. Boseggia for helpful discussions. This work is supported by the UK Engineering and Physical Sciences Research Council (Grants No. EP/N027671/1 and No. EP/N034694/1).

\end{acknowledgements}

\bibliography{Bibliography}

\end{document}